\documentclass[utf8]{FrontiersinHarvard} % for articles in journals using the Harvard Referencing Style (Author-Date), for Frontiers Reference Styles by Journal: https://zendesk.frontiersin.org/hc/en-us/articles/360017860337-Frontiers-Reference-Styles-by-Journal
\usepackage{url,hyperref,lineno,microtype,subcaption}
\usepackage[onehalfspacing]{setspace}
\usepackage{graphicx}
\usepackage{epstopdf}
\usepackage{tikz,siunitx}
\usetikzlibrary{calc, math, positioning}
\def\keyFont{\fontsize{8}{11}\helveticabold }
\def\firstAuthorLast{Hater {et~al.}} %use et al only if is more than 1 author
\def\Authors{Thorsten Hater\,$^{1,*}$, Juliette Courson\,$^{2,3}$, Han Lu\,$^{1}$, Sandra Diaz-Pier\,$^{1}$ and Thanos Manos\,$^{2,*}$}
\def\Address{$^{1}$Simulation and Data Lab Neuroscience, Jülich Supercomputing Centre (JSC), Forschungszentrum Jülich GmbH, Jülich, Germany \\
$^{2}$ETIS Lab, ENSEA, CNRS, UMR8051, CY Cergy-Paris University, Cergy, France \\
$^{3}$Department of Computer Science, University of Warwick, Coventry, UK}
\def\corrAuthor{Thorsten Hater}
\def\corrEmail{t.hater@fz-juelich.de}
\newcommand{\dd}[2]{\frac{\partial #1}{\partial #2}}

\begin{document}
\onecolumn
\firstpage{1}

%\title[Running Title]{Article Title} 
\title[Arbor-TVB co-simulator]{Arbor-TVB:\@ A Novel Multi-Scale Co-Simulation Framework with a Case Study on Neural-Level Seizure Generation and Whole-Brain Propagation}

\author[\firstAuthorLast]{\Authors} %This field will be automatically populated
\address{} %This field will be automatically populated
\correspondance{} %This field will be automatically populated

%\extraAuth{}% If there are more than 1 corresponding author, comment this line and uncomment the next one.
\extraAuth{Thanos Manos, ETIS Lab, ENSEA, CNRS, UMR8051, CY Cergy-Paris University, Cergy, France, thanos.manos@cyu.fr}

\maketitle

\section*{Article type: Methods}

\begin{abstract}
%%% Leave the Abstract empty if your article does not require one, please see the Summary Table for full details.
%\section{}
%For full guidelines regarding your manuscript please refer to \href{http://www.frontiersin.org/about/AuthorGuidelines}{Author Guidelines}.
Computational neuroscience has traditionally focused on isolated scales, limiting understanding of brain function across multiple levels. 
While microscopic models capture biophysical details of neurons, macroscopic models describe large-scale network dynamics. 
Integrating these scales, however, remains a significant challenge. 
In this study, we present a novel co-simulation framework that bridges these levels by integrating the neural simulator Arbor with The Virtual Brain (TVB) platform. 
Arbor enables detailed simulations from single-compartment neurons to populations of such cells, while TVB models whole-brain dynamics based on anatomical features and the mean neural activity of a brain region. 
By linking these simulators for the first time, we provide an example of how to model and investigate the onset of seizures in specific areas and their propagation to the whole brain. 
This framework employs an MPI intercommunicator for real-time bidirectional interaction, translating between discrete spikes from Arbor and continuous TVB activity.
Its fully modular design enables independent model selection for each scale, requiring minimal effort to translate activity across simulators.
The novel Arbor-TVB co-simulator allows replacement of TVB nodes with biologically realistic neuron populations, offering insights into seizure propagation and potential intervention strategies. 
The integration of Arbor and TVB marks a significant advancement in multi-scale modeling, providing a comprehensive computational framework for studying neural disorders and optimizing treatments.

\tiny
 \keyFont{ \section{Keywords:} Arbor, The Virtual Brain, Multi-scale Neural Models, Seizures, Mouse Brain Connectome} %All article types: you may provide up to 8 keywords; at least 5 are mandatory.
\end{abstract}

\section{Introduction}

The human brain consists of billions of neurons and an equally vast population of non-neuronal cells, intricately organized into layers and regions~\citep{herculano2009human,herculano2012remarkable}. Each neuron operates as a highly sophisticated biochemical machinery~\citep{west2002regulation,augustine2003local,darnell2013rna,Lu2025repetitive}, coordinating signal transmission within an extensive network in health~\citep{reyes2003synchrony,barral2019propagation,dicks2022gut} and disease [see e.g.,~\cite{tetzlaff2025characterizing}].
Ever since the Hodgkin-Huxley model was introduced to describe membrane potential dynamics~\citep{hodgkin1952quantitative}, computational neuroscience has played a pivotal role in enhancing our understanding of brain function. Yet, due to the immense complexity of the brain and computational constraints, most modeling studies focus on a single scale simulator or rely on standalone simulation codes. 

Modeling the large-scale electrical activity of the brain is a complex task. It
not only demands familiarity with advanced mathematical methods, but also a
solid grasp of the brain’s physiology and anatomy.. Neural field theory offers a
way to study the nonlinear behavior of large groups of neurons at a population
level, while still keeping the mathematics manageable. These models give us a
strong theoretical framework for understanding key processes in neural tissue,
including how the brain transitions between different activity states, such as
those seen in sleep or during epileptic events, see e.g.,~\citep{Cook_etal_2022}
for a recent review. Moreover, multi-scale computational modeling provides a
framework for connecting neural mechanisms with measurements ranging from unit
recordings to electroencephalogram (EEG), magnetoencephalography (MEG), and
functional magnetic resonance imaging (fMRI). Such models clarify how neural
systems compute and interact, and they are essential for integrating empirical
findings into a robust theoretical understanding of brain function, see e.g.,
\citep{Deco_etal_2008,Cooray_etal_2023}. Along this direction, recently in
\citep{Cooray_etal_2025}, the authors also studied oscillatory activity in
cortical tissue arising from not uniform neural connections and showed that
oscillations can be maintained under a wide range of anisotropic and
time-varying connectivity patterns.

Simulations incorporating biophysical properties and neural morphology typically
concentrate on individual neurons using simulators such as the NEURON simulator
\citep{TheNEURONBook}. At the microscopic level, studies have explored questions
such as how the ion channel kinetics influence neural excitability [see
e.g.,~\cite{gurkiewicz2011kinetic,suma2024ion}], how proteins, enzymes, and
calcium concentration are distributed among neighboring spines to impact
plasticity [see e.g.,~\cite{luboeinski2021memory,chater2024competitive}], and
how signal propagation along axonal fibers relates to neuropathic pain [see
e.g.,~\cite{tigerholm2014modeling,tigerholm2015c}]. Some studies examine how
neural morphology---such as dendritic tree growth [see
e.g.,~\cite{yasumatsu2008principles}] and morphology-dependent plastic
interactions [see e.g.,~\cite{hananeia2024multi}]---affects function. These
studies, while often limited to small patches of the neural membrane, a few
dendritic segments, or a small local network, provide valuable approximations of
broader neural phenomena.

At the mesoscopic level, researchers simplify neuronal representations using
point leaky-integrate-and-fire neurons (based on simulators such as NEST
\citep{Gewaltig:NEST} or Brian/Brian2~\citep{Stimberg2019}), allowing studies on
larger networks without explicit neuronal morphology or with some degree of
self-customized morphology, using, e.g., NESTML~\citep{linssen_2024_12191059}.
This approach has advanced our understanding of neural heterogeneity
\citep{demirtacs2019hierarchical,NEURIPS2021_656f0dbf,gast2024neural},
self-organization
\citep{zheng2013network,diaz2016automatic,miner2016plasticity}, neural capacity
\citep{emina2022selective}, energy efficiency~\citep{sacramento2015energy}, and
neural plasticity in disease and health~\citep{manos2021long,Lu_2024}. Most
microscopic and mesoscopic models remain theory-driven, using mathematical
approximations to infer neural behavior rather than directly establishing model
based on large datasets [see e.g.,~\cite{Popovych_2019} for a recent review].

Data-driven modeling has gained traction at the macroscopic level with the rise
of open-source brain imaging databases (such as OpenfMRI
\citep{poldrack2017openfmri}). High-resolution structural and functional data
from magnetic resonance imaging (MRI) and diffusion tensor imaging (DTI) enable
whole-brain modeling based on real anatomical features. The Virtual Brain (TVB)
\citep{sanz2013virtual,sanz2015mathematical,ritter2013virtual} and Virtual Brain
Twin (VBT)~\citep{Hashemi_etal_2025} platforms, for instance, integrate
functional MRI and DTI datasets to build individualized models, using coupled
oscillators to represent regional activity. TVB has contributed significantly to
the understanding of neurological disorders and serves as a testing ground for
therapeutic interventions [see
e.g.,~\cite{stefanovski2021bridging,monteverdi2023virtual,courson2024exploratory}
and references therein], for studying self-organization on macroscale [see e.g.,
\citep{Fousek2024}], consciousness [see e.g.,~\citep{Breyton_etal_2024} and
healthy aging [see e.g.,~\citep{Lavanga_etal_2023}].

With advances in computing resources and simulation technologies, the integration of models across different scales has become both feasible and essential to strike a balance between retaining detailed information and achieving a broad-scale understanding. Recently, a co-simulation framework was introduced that employs NEST and TVB to bridge mesoscopic and macroscopic modeling. This work has pioneered cross-scale modeling~\citep{kusch2024multiscale} and has demonstrated the benefits of integrating models across spatial levels. A notable application of the virtual deep brain stimulation model~\citep{Meier2022,Shaheen2022,Wang_etal_2025} demonstrated its utility in multiscale simulations. Similar tools have been made available within the European digital neuroscience platform, EBRAINS~\citep{Schirner2022}. 

However, integrating microscopic and macroscopic models remains technically challenging. 
At the core lies the vast amount of information being processed, billions of cells with thousands of connections, and the immense gap in timescales, from microseconds in ion channel dynamics to minutes or hours for plastic changes of the connectome.
To solve this challenge, we used the Arbor simulator~\citep{arbor2019} and, more specifically, its most recent next-generation version~\citep{cumming_2024_13284789}, at the microscopic end. Designed for single-neuron and large-scale network simulations, Arbor leverages GPU resources to enhance computational speed and energy efficiency.

In this Methods paper, we successfully established efficient communication between Arbor and the TVB that respects their different operational time scales and provided a use case example of the cross-scale interaction. 
To demonstrate a first showcase, we used a mouse brain connectome provided by TVB, where each region represents the mean mass neural activity of a brain area modeled by a macroscopic model. 
Using the co-simulation interface, we replaced one TVB node with a network of detailed neurons modeled in Arbor.
An extended Hodgkin-Huxley-based neuron model~\citep{depannemaecker2022unified} was utilized in Arbor to simulate different neural activity patterns (e.g., spiking, bursting, seizure-like, etc.).
By tweaking a single parameter, we showcased that the seizure-like events generated in Arbor propagated to other nodes modeled in TVB.\@ 
This platform provides users with the freedom to use existing models across scales with minimal additions and enabled future development in building brain digital twins that contain both micro- and macroscopic information for therapeutic applications.

\section{Materials and methods}

\subsection{The Arbor simulator}%
\label{sec:arbor}
 
Arbor is an open-source library for building simulations of biophysically detailed neuron models~\citep{arbor2019}. It provides an alternative to software like NEURON~\citep{TheNEURONBook}, but with a strong emphasis on modern hardware and scalability to large-scale systems~\citep{hines1984}. Its overall set of capabilities allows Arbor to model neural networks at a level of resolution beyond point models to explore phenomena like dendritic computation.
Thus, support for bulk-synchronous parallelism via MPI, shared memory parallelism by utilizing a thread-pool and job system is central to Arbor, and certain cell types---primarily cable cells---can further leverage SIMD and GPU hardware. Arbor is written in $C$++, though most users interface with it through an intuitive, high-level Python interface built on top of the lower level implementation.
The underlying numerical model of Arbor is the cable equation:
\begin{align}
    c \dd{U}{t} = \dd{}{x}\left(\sigma\dd{U}{x}\right) + i
    \label{eq:cable}
\end{align}
where the membrane potential $U$ is computed over the morphological structure of the neural tree; the spatial coordinate $x$ and the derivative are to be understood within this structure~\citep{von1850messungen,thompson1855theory,hodgkin1952measurement,hodgkin1952currents,hodgkin1952components,loligo1952dual,hodgkin1952quantitative,scott1975electrophysics}.
The parameters $c$ and $\sigma$ define the membrane capacitance and longitudinal resistance. The trans-membrane current density $i$ models the entirety of ionic and non-ionic currents.
In both NEURON and Arbor, these are calculated from user-specified sets of differential equations, potentially varying along the morphology.
The equations for $i$ and $U$ are solved in alternation (Lie-Trotter splitting) using a first-order implicit method.

\begin{figure}[t]
    \centering
    \includegraphics[width=\textwidth]{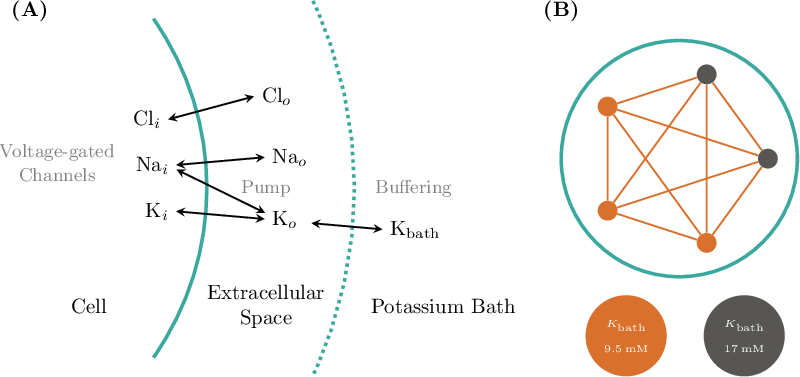}
    \caption{%
    \textbf{Biophysical neuron model and Arbor network.} 
    \textbf{(A)} For single cell dynamics three ion concentrations (K, Na, and Cl) are modeled in the cell's interior and a thin shell of its extracellular medium. The latter is, in turn, surrounded by a bath of a fixed potassium concentration $\mathrm{K}_\mathrm{bath}$. The model simulates changes to the concentration in addition to the current contributions based on three voltage-gated ion channels, an active pump between potassium and sodium, and the buffering effect of the surrounding potassium bath.
    \textbf{(B)} We choose typical values of $K_\mathrm{bath}$ for the single models to generate the tonic spiking and seizure-like event behaviors.
    In most cases, a fully connected network using exponential synapses with weight $w=0.5$ is used. 
    As an example, we show here the network instantiation for a network size $N=5$ and a ratio of SLE to tonic neurons of $f=0.2$. 
    }%
    \label{fig:arbor-model}
\end{figure} 

%\hfill \break

\subsection{Single neuron and network models in Arbor}
\begin{figure}[t]
    \centering
    \includegraphics[width=1.0\textwidth]{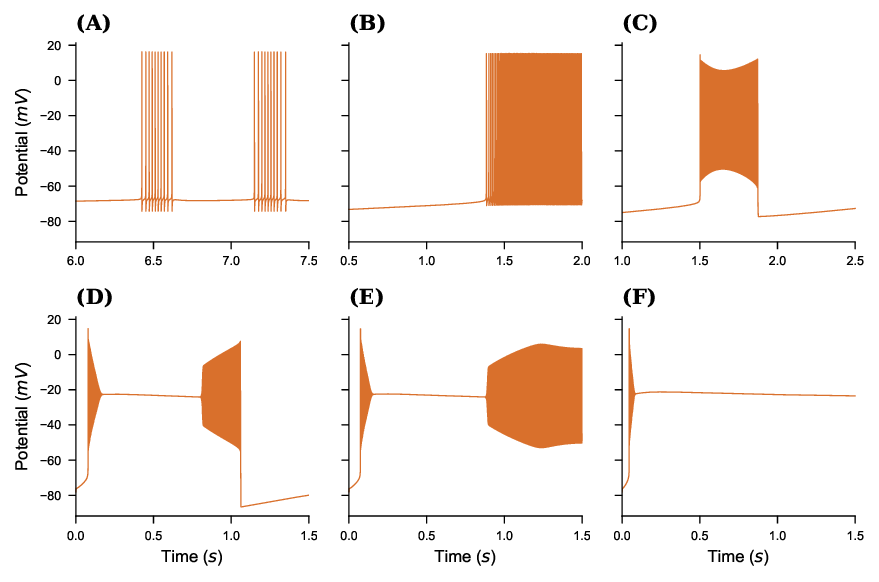}
    \caption{\textbf{Different neural spiking patterns.}
    \textbf{(A)} Spike Train, $\mathrm{K}_\mathrm{bath}=\qty{7.5}{mM}$. \textbf{(B)} Tonic Spikes, $\mathrm{K}_\mathrm{bath}=\qty{9.5}{mM}$. \textbf{(C)} Bursting, $\mathrm{K}_\mathrm{bath}=\qty{12.5}{mM}$. \textbf{(D)} Seizure-Like Event (SLE), $\mathrm{K}_\mathrm{bath}=\qty{17.0}{mM}$. \textbf{(E)} Sustained Ictal Activity (SIA), $\mathrm{K}_\mathrm{bath}=\qty{17.5}{mM}$. \textbf{(F)} Depolarization Block, $\mathrm{K}_\mathrm{bath}=\qty{22.5}{mM}$. Note that by setting $\mathrm{K}_\mathrm{bath}=\qty{4}{mM}$, one can obtain Resting State (RS) activity too (result not shown here). 
    }%
    \label{fig:single_neuron_kbath}
\end{figure}

We begin by selecting a dynamical model that allows for relatively easy yet realistic simulation of a broad spectrum of neural activity at the single-neuron level, governed by a small set of biophysical parameters. 
The neural model from~\citep{depannemaecker2022unified} was formulated for Arbor in the Neuron MODeling Language (NMODL). 
The following equations form the slow part of the system, describing the evolution of ion concentrations due to voltage-gated channels, active pumps, and buffering by an external bath, see \textbf{Figure}~\ref{fig:arbor-model}(A) for a schematic of the dynamics. 
It describes the ionic exchanges between the intracellular and extracellular spaces (ICS, ECS) of a neuron immersed within an external
bath, acting as a potassium buffer of concentration $\mathrm{K}_\mathrm{bath}$.
Ions flow between the ICS and ECS through a sodium-potassium pump and the sodium, potassium and chloride voltage-gated channels, driving changes in the internal ($\mathrm{K}_i$, $\mathrm{Na}_i$, $\mathrm{Cl}_i$) and external ($\mathrm{K}_o$, $\mathrm{Na}_o$, $\mathrm{Cl}_o$) ionic concentrations.
By gradually increasing the external bath concentration of potassium ions $\mathrm{K}_\mathrm{bath}$, the model sequentially presents these patterns: resting state (RS), spike train (ST), tonic spiking (TS), bursting, seizure-like events (SLE), sustained ictal activity (SIA) and depolarization block (DB), see \textbf{Figure}~\ref{fig:single_neuron_kbath}. 
The fast dynamics of the membrane potential $V$ is modeled in Arbor via the cable equations, see above, which require computing the ion current densities $i_{\mathrm{X}}$ used in the simulator update as:
\begin{align}
    i_{\mathrm{X}} &= g_{\mathrm{X}}(V - E_{\mathrm{X}})\\ 
    E_{\mathrm{X}} &= C\cdot\log\left(\frac{\mathrm{X}_o}{\mathrm{X}_i}\right)
\end{align}
with the ion species $\mathrm{X} = \left\{\mathrm{K}, \mathrm{Na}, \mathrm{Cl}\right\}$
and a non-ion current density:
\begin{align}
    i_\mathrm{pump} = \frac{\rho}{(1 + \exp(10.5 - 0.5\mathrm{Na}_i))(1 + \exp(5.5 - \mathrm{K}_o))}.
\end{align}
These currents enter the cable equation Eq.~\eqref{eq:cable} as the trans-membrane current $i$ via:
\begin{align*}
    i = i_\mathrm{pump} + \sum_X i_X
\end{align*}
Following the Hodgkin-Huxley model, conductivities $g_{\mathrm{X}}$ are written as:
\begin{align}
\begin{split}
    g_{\mathrm{K}} &= g_{0, \mathrm{K}}n + g_{l, \mathrm{K}}
\end{split}
\quad
\begin{split}    
    g_{\mathrm{Na}} &= g_{0, \mathrm{Na}}m h + g_{l, \mathrm{Na}}
\end{split}
\qquad\qquad
\begin{split}    
    g_{\mathrm{Cl}} &= g_{0, \mathrm{Cl}}
\end{split}
\end{align}
The --- internal $i$ and external $o$ --- ion concentrations are modeled as:
\begin{align}
\begin{split}
    \mathrm{K}_i &= \mathrm{K}_{0, i} + \Delta\mathrm{K}_i\\
    \mathrm{K}_o &= \mathrm{K}_{0, o} - \beta\Delta\mathrm{K}_i + \mathrm{K}_g
\end{split}
\quad
\begin{split}
    \mathrm{Na}_i &= \mathrm{Na}_{0, i} - \Delta\mathrm{K}_i\\
    \mathrm{Na}_o &= \mathrm{Na}_{0, o} + \beta\Delta\mathrm{K}_i
\end{split}
\qquad\qquad
\begin{split}    
    \mathrm{Cl}_i &= \mathrm{Cl}_{0, i}\\
    \mathrm{Cl}_o &= \mathrm{Cl}_{0, o}
\end{split}    
\end{align}
The variables $\left\{\Delta\mathrm{K}_i, \mathrm{K}_g\right\}$ evolve as:
\begin{align}
    \frac{d\Delta\mathrm{K}_i}{dt} &= -\gamma\left(i_\mathrm{K} - i_\mathrm{pump}\right)\\
    \frac{d\mathrm{K}_g}{dt} &= \epsilon\left(\mathrm{K}_\mathrm{bath} - \mathrm{K}_o\right)
\end{align}
where $\gamma$ converts the current density $i_\mathrm{X}$ to molar flux, summarizing the effect of the ion pump in \textbf{Figure}~\ref{fig:arbor-model}(A) and the external buffer. Finally, fast dynamics were reduced and adjusted to mammalian neurons:
\begin{align}
    \frac{dn}{dt} &= \frac{1}{\tau}\left(n - n_\infty(V)\right)\\
    n_\infty(V) &= \frac{1}{1 + \exp(-(19 + V)/18)}\\
    m = m_\infty(V) &= \frac{1}{1 + \exp(-(2 + V/12))}\\
    h = h(n)        &= 1.1 - \frac{1}{1 + \exp(3.2 - 0.8n)}
\end{align}
based on the observations that the reaction of the sodium gating variable to changes in $V$ is nigh instantaneous and $h(t) + n(t) = \mathrm{const}$.

The resulting ion channel was added to a basic, spherical, single-compartment neuron. After implementing this biophysical model, we reproduced the firing patterns using the parameters of the reported model [\textbf{Figure}~\ref{fig:single_neuron_kbath}~(A-F)], see also~\citet{depannemaecker2022unified} for more details and motivation for model parameter choices. Note that despite the values given in the original publication, neither the Arbor nor the published reference implementation produces the depolarization block pattern at $K_\mathrm{bath} = \qty{20}{mM}$ but only at around $K_\mathrm{bath} = \qty{22.5}{mM}$. 
From here, a simple model network was developed, comprising $N$ total cells, with a mixture of tonic $f\cdot N$ and SLE $(1-f)\cdot N$ cells, where both sub-populations are assigned individual values for $K_\mathrm{bath}$, sketched in \textbf{Figure}~\ref{fig:arbor-model}. 
Cells are connected using exponential synapses with an internal weight of $w=0.5$ chosen to produce an activity similar to~\cite{Rabuffo_etal_2025} which uses delta synapses.

\begin{figure}
    \centering
    \textbf{(A)}
    \includegraphics[width=0.11\linewidth]{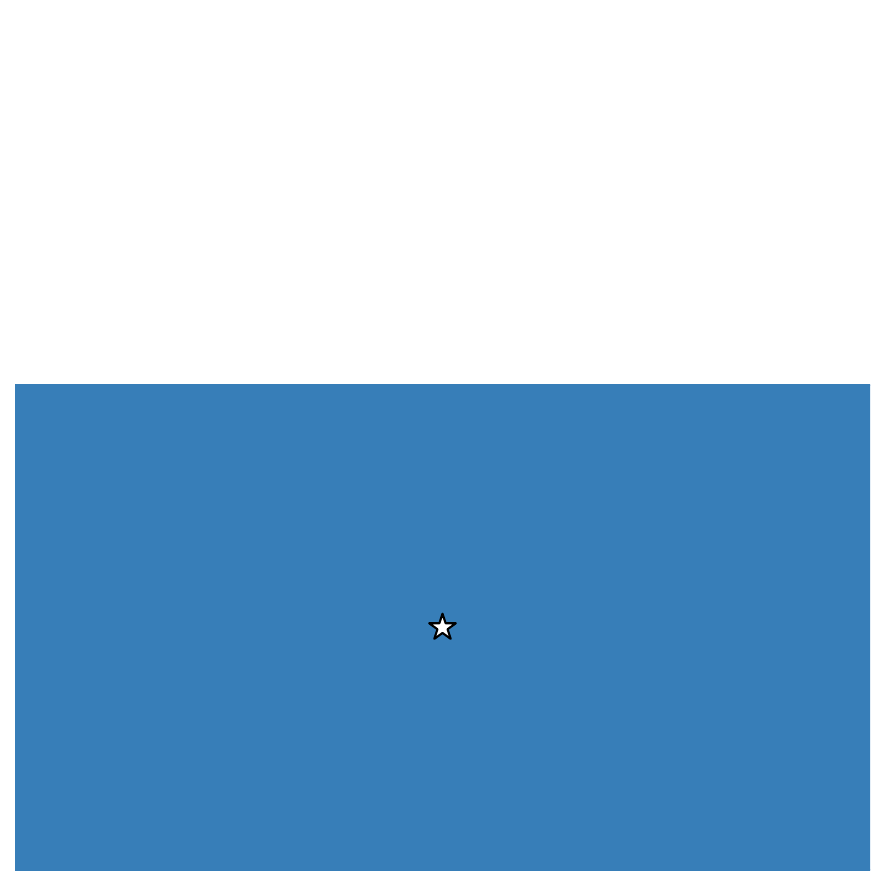}
    \qquad
    \textbf{(B)}
    \includegraphics[width=0.2\linewidth]{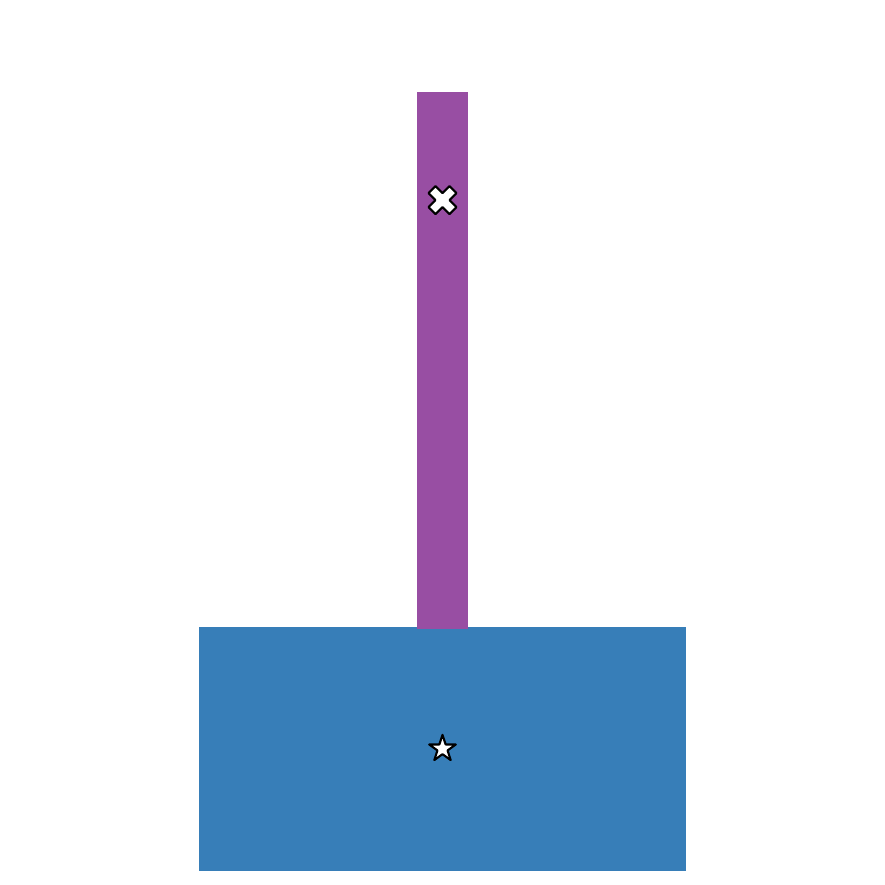}
    \textbf{(C)}
    \includegraphics[width=0.3\linewidth]{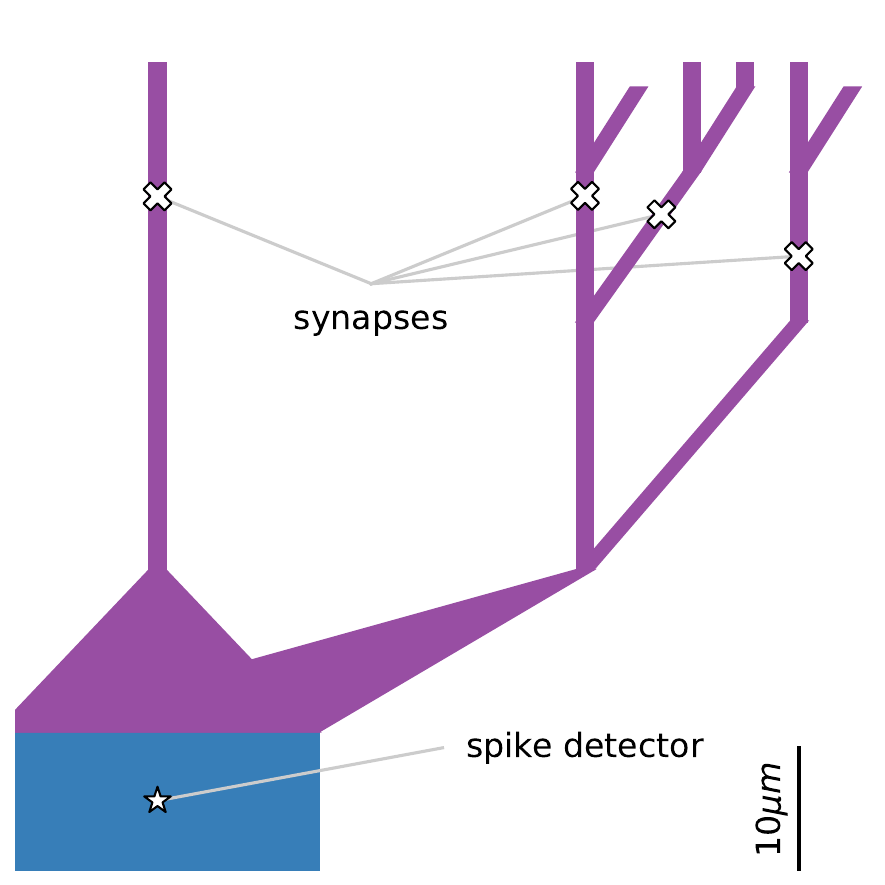}
    \caption{\textbf{Compartmental neuronal morphologies used for simulations.}
      Each morphology comprises a soma (blue) and a dendritic section (violet).
      For numerical simulations, dendritic segments are discretized into
      \qty{5}{\mu m} compartments, whereas the soma is modeled as a single
      compartment. Since the cable model neglects extracellular effects, a
      $1.5$-dimensional morphology is employed internally, rendering spatial
      relationships irrelevant to the model dynamics. When multiple connections
      converge onto a single cell, synaptic assignments follow a round-robin
      scheme. \textbf{(A)} Soma only. Equivalent to a point model, the
      cylindrical segment has a radius $r$ and a length of $2r$, chosen to yield
      the same surface area as a sphere of radius $r$. Both synapses and spike
      detectors are attached at the center of the soma. \textbf{(B)} Ball and
      stick. A straight dendritic segment is added, featuring passive current
      flow and a synapse attached at a fixed distance from the soma.
      \textbf{(C)} Random trees. More complex morphologies are generated as
      random binary trees of depth five. Synapses are placed at a fixed distance
      from the soma and may be targeted by connections originating from spike
      detectors, which are positioned at the soma center. See \textbf{Figure}~S1
      in the Supplemental Material for more examples of randomly generated
      morphologies.}%
    \label{fig:single-cell-trees}
\end{figure}

In addition to the elementary spherical morphology, we also investigated two multi-compartmental neuron models. The first model included a single dendritic segment of \qty{25}{\mu m}, subdivided into \qty{5}{\mu m} compartments. The second model extended the dendrite into a randomly generated tree composed of \qty{5}{\mu m} compartments. To introduce variability across cells, the random number generator was seeded with each cell’s unique identifier (see \textbf{Figure}~\ref{fig:single-cell-trees} for examples). In future work, these synthetic morphologies will be replaced with reconstructions derived from neural imaging data available in public databases. Supporting such models in Arbor will require only a simple command to load per-cell morphology data from disk.

\subsection{TVB network model}
Following~\citet{Deco_2013}, we used the reduced Wong-Wang model~\citep{Wang_2002} to simulate resting-state activity and to investigate the dynamics of local brain regions embedded within a large-scale brain network. The mean firing rate $H(x_I)$ and mean synaptic gating variable $S_I$ of region $I$ are described by:
\begin{align}
%    \dd{S_I}{t} &= -\frac{S_I}{\tau_s} + (1-S_I)\gamma H(x_I)\\
    \frac{dS_I}{dt} &= -\frac{S_I}{\tau_s} + (1-S_I)\gamma H(x_I)\\
    H(x_I) &= \frac{ax_I - b}{1-\text{exp}(-d(ax_I -b))} \\
    x_I &= \omega J_N S_I + G J_N \sum_K c_{IK}S_K + I_0,
\end{align}
where $x_I$ is the synaptic input to the $I$-th region, $\omega=1$ denotes the local excitatory recurrence, $c_{IK}$ is the strength of the structural connection from the local area $I$ to $K$, and $G$ is a global coupling strength. The parameters are set to the same values as those used in the TVB implementation. $J_N=\qty{0.2609}{nA}$ is the synaptic coupling of NMDA receptors and $I_0=\qty{0.33}{nA}$ is the baseline external input. The kinetic parameters are $\tau_s=\qty{100}{ms}$ and $\gamma=0.641$. The parameters of the input-output function $H$ are $a=\qty{0.27}{nC^{-1}}$, $b=\qty{0.108}{kHz}$, and $d=\qty{154}{ms}$. Depending on the tuning of $G$, the system exhibits a multi-stable regime, with steady states of high and low spiking activity. Here, we set $G=0.096$.

\subsection{Co-simulation framework of Arbor and TVB}

\begin{figure}[t]
    \centering
    \hspace*{-1.0cm}  
    \includegraphics[width=\textwidth]{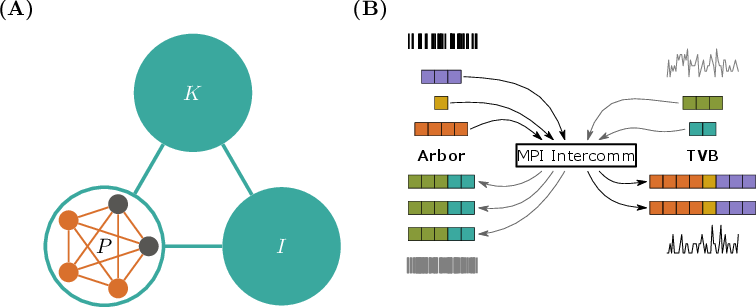}
    \caption{ \textbf{Arbor-TVB co-simulation schematic and communication
        pattern.} \textbf{(A)} In a TVB simulation of regions $I$, $K$, and $P$,
      one region $P$ will be replaced by a proxy containing a network of
      detailed cells simulated in Arbor. Regions are connected via the weights
      of the connectome and produce an activity values based on the chosen
      region model. When crossing the boundary between TVB and Arbor models,
      care needs to be taken to convert between discrete action potentials in
      Arbor to continuous, region-model-specific variables in TVB.
      \textbf{(B)} Spikes generated by Arbor and TVB --- converted from activity
      values interpreted as mean spiking rates --- are exchanged using an MPI
      intercommunicator and the All-gather primitive. This is equivalent to
      concatenating all contributions from all Arbor MPI ranks and sending the
      result to all TVB ranks and vice-versa.}%
    \label{fig:tvb-model}
\end{figure}

Both Arbor and TVB offer support for attaching a second simulator to perform co-simulation, potentially at different scales. 
% TVB
Co-simulation from TVB's viewpoint is the simpler technology of the two frameworks, since TVB is designed to execute as a single process. TVB allows for exchange of any variable relevant to the region models and any number of variables. 
The co-simulation partner is encapsulated in one or more TVB regions, called proxy nodes, see \textbf{Figure}~\ref{fig:tvb-model}(A). 
These proxies present a conforming interface to TVB, exchanging the salient variables as a table, one row per time-step, one column per variable. 
As TVB advances in lockstep on a global time-step, this is almost identical to normal operation. 
However, co-simulation introduces the concept of an \emph{`epoch'} to TVB, i.e., the length of time that conforms to the smallest delay $\tau_\mathrm{min}$ in the set of inter-region connections delays ${\tau_{IJ}}$, with $I$ and $J$ referring to two connected regions. These delays are part of the connectome data used to construct a TVB simulation. 
In the case that a connectome contains zero-valued delays these must be replaced with a pre-defined finite value. 
Further, it is required that the time-step evenly divides $\tau_\mathrm{min}$. Co-simulation thus can integrate all nodes' state, including the proxy, for one epoch $\tau_\mathrm{min}$ without exchanging data. This is correct as an event emanating from any region $I$ at time $t$ influences any other region $J$ at time $t + \tau_{IJ} \geq t + \tau_\mathrm{min}$. Only after an epoch, data need to be exchanged between the proxy and the rest of the TVB regions. A TVB---NEST demonstration has been published to showcase the interaction between a local network of spiking neurons and the whole-brain network dynamics~\citep{kusch2024multiscale}.

\begin{figure}[t]
    \centering
    \includegraphics[width=1.0\textwidth]{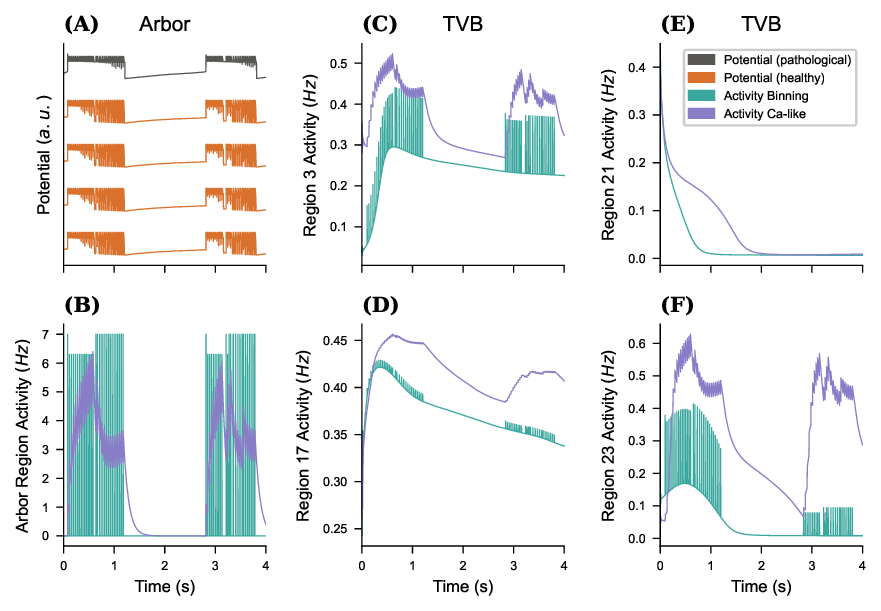}
    \caption{\textbf{Impact of conversion method on activity exchange.} For an
      all-to-all connected network of a mixture of \num{10} SLE cells
      ($K_\mathrm{bath} = \qty{17.5}{mM}$) and \num{90} tonic
      ($K_\mathrm{bath} = \qty{9.5}{mM}$) neurons in Arbor, we plot the membrane
      potential traces for four tonic and one SLE cell in \textbf{(A)}. This
      simulation is repeated for two activity exchanging methods, either spikes
      were binned into buckets of width $\Delta t$ to extract instantaneous
      rates, or the differential equation Eq.~\eqref{eq:calcium} emulating the
      change in Calcium concentration of a biological cell after spiking was
      used (with $\tau = \qty{100}{ms}$ and $\beta = \frac{0.1}{N}$). The
      resulting activity traces for the Arbor network \textbf{(B)} and selected
      TVB nodes (out of 98 regions) are displayed in panels \textbf{(C---F)}.}%
    \label{fig:co-sim-example}
\end{figure}

% Arbor
Arbor has a different design in terms of connectivity.
Interaction between physically separate cells is mediated by action potentials, e.g., when the membrane potential triggered by dedicated sources crosses a configurable threshold. Cells are connected by wiring these sources to corresponding sinks like synapses via an abstract connection object comprising a delay and weight, modeling transmission and attenuation via an axon. 
In contrast to TVB, Arbor is fundamentally a distributed system and internally employs the same approach to decoupling via the minimum network delay as explained above. To initiate co-simulation, Arbor utilizes an additional interface to manage the spike exchange from external connections that are originated from outside but terminate at cells simulated in Arbor.
On the technical side, the latter part leverages \verb!MPI_Allgatherv! through an inter-communicator and effects that the concatenation of all spikes sent from all MPI ranks running TVB arrive on all ranks running Arbor and vice versa [\textbf{Figure}~\ref{fig:tvb-model}(B)]. 
This allows co-simulation in conjunction with arbitrary numbers of ranks on both sides and even in compounds with more than two simulators.

% bringing Arbor and TVB together
Finally, bi-directional translation between TVB's variable concept and Arbor's representation of action potentials is required. As the former depends on the region models used, we chose to bundle this with the remaining TVB functionality as part of the Arbor proxy node.
% rate -> spike
For the TVB models used in this study, the main variable is the per-region mean activity rate $\nu_I$ which is conceptually compatible with the concept of spike generation. For each region $I$ connected to the proxy node $P$, i.e., with connectome weight $c_{IP} > 0$, a set of synthetic events must be generated such that the mean activity conforms to $\nu_I$.
This is an ambiguous process, even if we prescribe a population (list of cell identifiers) and a per-cell distribution, e.g., a Poisson point process, from which to draw events, which likely must be resolved by ensembles of simulations. In general, this is both model dependent and mathematically intractable, so we leave the general case as a customization point in the framework.  
For our running example, however, we make the following choice: 
Event timings for the current step $k$ will be drawn from a uniform distribution and dispatched to all cells in the Arbor network.
Note that while these events are created at given time a per-connection delay is applied and thus delivery occurs at a later time.

% spike -> rate
The inverse direction, converting spike events to mean rates, while being well-defined, is still subject to customization. We explore two options here. 
First, simple running averages, i.e., all spikes that originate within the Arbor network during the current epoch, are collected and sorted into bins of width $\Delta t$. 
This list is then normalized to the cell count and time step and sent to TVB as the mean activities as a function of time.
Although straightforward, this can lead to unrealistically rough activity traces, especially if cell populations are small. 
Second, as inspired by high-speed calcium imaging experiments~\citep{grewe2010high}, a mechanism to track cell activity through calcium level is implemented as:
\begin{align}
   \frac{dC_p}{dt}  = - \frac{C_p(t)}{\tau} + \beta\sum_{t_{\mathrm{spike}, p}}\delta(t - t_{\mathrm{spike}, p}),\qquad C_p(0) = 0
   \label{eq:calcium}
\end{align}
with per cell $p$ having a decay parameter $\tau$ and a weight $\beta$. Computing the activity becomes the average:
\begin{align}
\nu_P(t) = \left\langle C_p(t)\right\rangle_{p \in P},
\end{align}
yielding a smooth trace. 
This method of converting discrete spiking events into a continuous interval variable is also used in a few plasticity models recruiting a negative feedback control mechanism such as synaptic scaling~\citep{van2000stable} and homeostatic structural plasticity~\citep{butz2013simple,diaz2016automatic,Lu_2024} models. The choice of the calcium kernel parameters is based on trial-and-error to match the output of Arbor and the magnitude of activity generated by TVB.\@ \textbf{Figure}~\ref{fig:co-sim-example} compares the impact of this choice on the macro-scale network. 
In small networks and over short timescales defined by the epoch length as shown in the example, spiking activity occurs in noncontinuous bursts, which is dubious in conjunction with the smooth dynamics of the chosen TVB model.
\textbf{Figure}~\ref{fig:co-sim-example}(A) shows the propagation of this noncontinuous activity into the TVB regions, while using the Ca-like model (B) provides smooth dynamics in both the Arbor and TVB models. 
We thus will use the latter in all simulations from here on out. 
In general, both methods require scaling by the number of cells in the proxy region to arrive at a scale-free activity measure. 
A local scaling factor $G_\mathrm{A}$ is used to convert between the activity of the detailed network and the activity of the region modeled in TVB.\@
In general, $G_\mathrm{A}$ needs to be adjusted to the choice of connectome and TVB model, similar to the choice of the global coupling strength $G$ in the RWW model. In this study, $G_\mathrm{A}=100$ is used as it produces seizure-like propagation patterns comparable to those found in similar studies, see e.g.,~\citet{Melozzi_etal_2017,courson2024exploratory} and references therein.

\section{Results}
So far, we have described the components of the co-simulation framework.
It consists of the following components: a TVB model based on the connectome and node dynamics, a specified set of TVB nodes where the Arbor models will be placed, one or more internally connected network models in Arbor, a defined mechanism for routing events from TVB to individual cells in Arbor, a method for translating Arbor-generated events into TVB variables, and a translation process for converting TVB variables into events originating from synthetic cells. Each of these components serves as a customization point for the user. While reasonable default configurations can be provided for some, others require user-defined specifications to suit specific modeling needs.

The single cell model has been demonstrated to exhibit the necessary range of behaviors.
We have also motivated our choice for converting spikes to rates of using a biologically-inspired exponential smoothing filter via a Ca-like activity over simple binning and fix parameters to $\tau=\qty{100}{ms}$ and $\beta=\frac{0.1}{N}$.
This normalization is important, as it produces results that are invariant under changes in the number of cells $N$ in the detailed network.

To induce seizure propagation in mice brain, we use the structural connectivity derived from the Allen mouse brain atlas~\citep{Oh_2014}, also used in~\cite{Melozzi_etal_2017}, to embed the TVB nodes. Following the work presented in~\cite{courson2024exploratory}, the proxy node modeling the Arbor population of tonic-spiking and SLE point neurons is set within left the Hippocampus and in particular in the left-field CA1 (l~CA1), a region that is prone to generate widespread seizures.

\subsection{Seizure induction among morphologically detailed cells}

\begin{figure}[t]
    \centering
    \begin{tikzpicture}
    \node at (0, 0) {\includegraphics[width=1.0\textwidth]{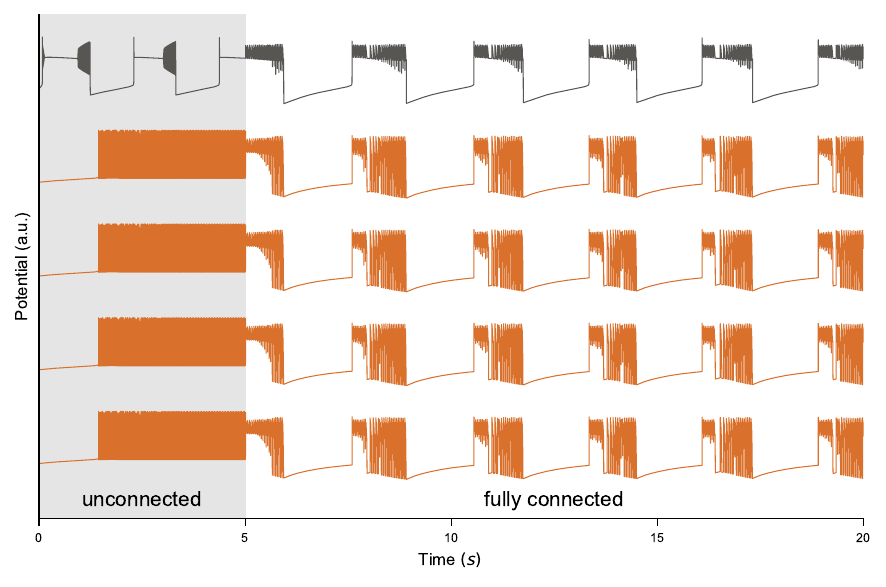}};
    \node[draw=black!80, fill=white] at (7, -2.5) {\includegraphics[height=3cm]{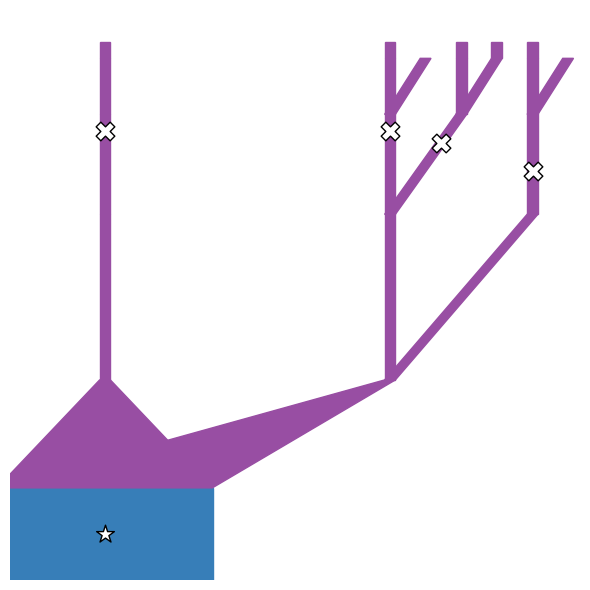}};
    \end{tikzpicture}
    \caption{\textbf{Network level effects induced by SLE activity.}
    Simulation of a 100-cell network with 80 tonic-spiking ($K_\mathrm{bath}=\qty{9.5}{mM}$) and 20 SLE ($K_\mathrm{bath}=\qty{17.0}{mM}$) neurons. 
    Membrane potentials are shown for one SLE (top) and four tonic-spiking cells.
    Each neuron includes a single-compartment soma and a random dendritic tree. 
    The model was integrated for \qty{5}{s} (shaded region) without internal connections, then switched to a fully connected network, which settled into a new equilibrium dominated by SLE activity. 
    Inset: Example morphology of the first cell.
    }%
    \label{fig:recruiting}
\end{figure}

As an initial step, we examined a network of morphologically detailed cells
(comprising of a single-compartment soma and a random dendritic tree, see
bottom-right inset panel of \textbf{Figure}~\ref{fig:recruiting}) without
embedding them into a co-simulation framework. The objective was to assess how
interactions among neurons exhibiting distinct firing patterns influence network
dynamics within a small population of SLE neurons at the local (Arbor) level. We
constructed a network comprising 80 tonic-spiking neurons
($K_{\text{bath}} = \qty{9.5}{mM}$) and 20 SLE neurons
($K_{\text{bath}} = \qty{17.0}{mM}$), initially without internal synaptic
connectivity. The system was simulated for $T = \qty{5}{s}$, and the resulting
membrane potentials are presented in \textbf{Figure}~\ref{fig:recruiting} (this
initial phase is indicated by the gray-shaded region). During this initial
phase, all neurons independently showed their intrinsic firing behavior,
consistent with the activity shown in
\textbf{Figure}~\ref{fig:single_neuron_kbath}. Following this baseline
simulation, integration was paused, the network was reconfigured to full
connectivity, and the simulation was resumed. The introduction of connectivity
produced an immediate response across the network. The SLE neurons progressively
transitioned toward tonic-like spiking, mirroring the dominant (80\%)
tonic-spiking subpopulation. At the same time, tonic-spiking neurons began to
exhibit the bursting activity characteristic of the SLE phenotype
(\textbf{Figure}~\ref{fig:recruiting}, $T \geq \qty{5}{s}$). Note that the use
of different $K_{\text{bath}}$ values for neurons within the same group is, at
this stage, driven by computational considerations and the need to validate the
performance of the Arbor-TVB co-simulator. This choice can be adjusted to study
more biologically realistic scenarios which involve the interaction of different
neural populations or transition of network dynamics, which is further addressed
in the Discussion section.

\subsection{Seizure induction and  propagation employing point neurons in Arbor-TVB co-simulator}

\begin{figure}[htbp]
    \centering
    \includegraphics[width=1.0\textwidth]{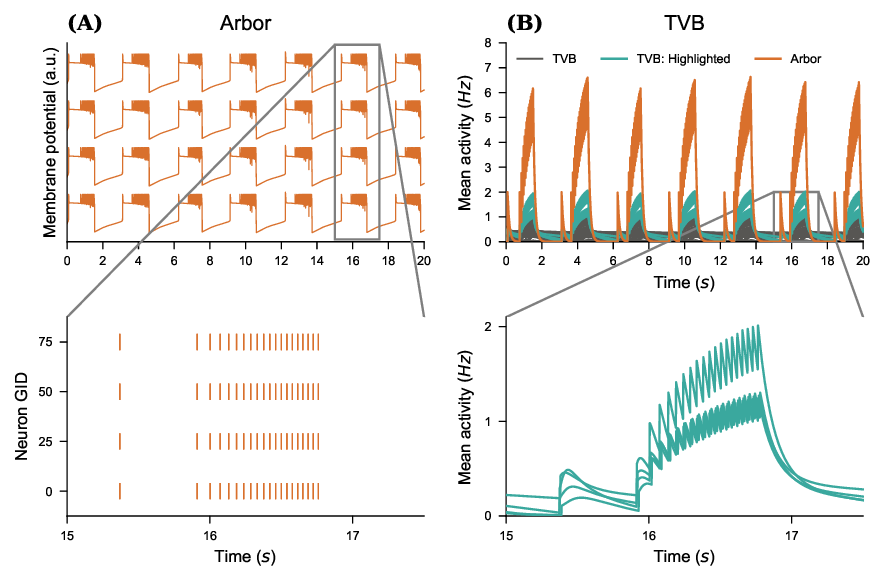}
    \caption{\textbf{Multiscale seizure propagation.} \textbf{(A)} Membrane
      potential and raster plot for four neurons in the detailed fully-connected
      Arbor neural network. Note that the membrane potential time-series are
      similar across all neurons. \textbf{(B)} Mean firing rate of various TVB
      local brain areas versus the Arbor activity. We track a seizure after a
      transient period, so that all brain areas have reached their baseline
      activity. The zoomed-in sections show the propagation of the seizure. We
      highlight and show in detail the four time series with the largest
      deviation in activity.}%
    \label{fig:tseries_cosim}
\end{figure}

Next, we embedded a network of detailed neurons as a proxy node in a simulation
of neural mass models in TVB.\@ The general setup is similar as before, however,
the proxy node now consists of only SLE-type point neurons in a fully connected
network. \textbf{Figure}~\ref{fig:tseries_cosim}~(A) illustrates the evolution
of the membrane potential of individual neurons within the Arbor network, with
all neurons exhibiting identical dynamical behavior. The pattern of SLE activity
is modulated by neuronal coupling.

We run simulations over \qty{20}{s}, and investigate the effect of SLE emergence in the network once all brain areas have reached their steady state. The production of these patterns in the Arbor node generates changes in the firing rates of local TVB brain areas. Even though changes in activity occur in most brain areas, these fluctuations occur within different ranges.

In \textbf{Figure}~\ref{fig:tseries_cosim}(B), we show the time-series of TVB nodes' firing rate throughout the simulation, here with $N=\num{100}$ in the Arbor detailed network. 
A transient period is necessary before all nodes reach their steady-state. 
The emergence of recurrent SLE patterns in the Arbor node triggers a mixture of periodic and seizure-like patterns in the dynamics of the TVB nodes.
We highlight traces corresponding to the l~CA1 region modeled using detailed cells (orange) and the four regions with the highest activity modeled by the neural mass model.
The inset shows the highlighted traces, excluding l~CA1, during a single period.

\begin{figure}[htbp]
    \centering

    \includegraphics[width=\textwidth]{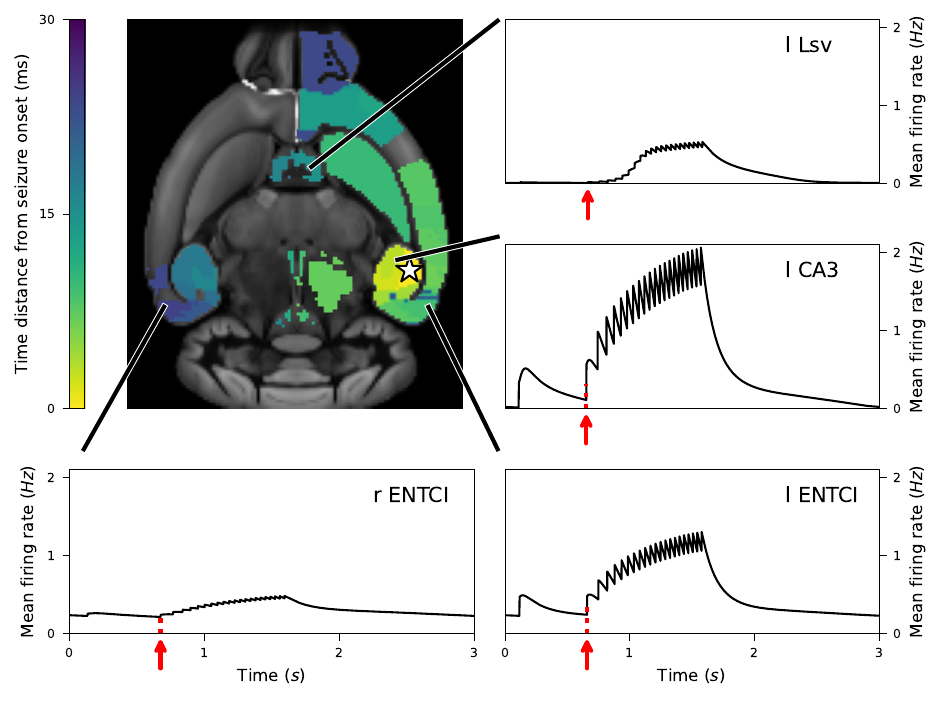}
    \caption{\textbf{Propagation of a seizure originating in left-field l~CA1
        area of the mouse brain model.} Time distance between seizure emergence
      in l~CA1 (star marker) and spiking rate increase in each brain area. We
      also depict the firing activity time-series of four initially non-SLE
      nodes of the brain network being recruited in the seizure, namely the
      ventral part of left Lateral Septal Nucleus (l~LSv), l~CA3, l~ENTCl and
      r~ENTCl. The red arrows indicate the beginning of the bursting activity.
      Non-colored regions (panel with the mouse brain template) correspond to
      areas where seizures either did not occur or had relatively weak effects.
      See text for more information regarding the onset detection of seizures in
      other regions. See text for more details.}%
    \label{fig:seizure_prop_ca1}
\end{figure}

In \textbf{Figure}~\ref{fig:seizure_prop_ca1}, we present the propagation of SLE originating in l~CA1 (star marker), represented as a fully connected network of $N=\num{10000}$ SLE point neurons.
Colors on the brain template show the time distance between seizure emergence in the diseased area and seizure arrival in the different brain regions. The systematic detection of SLE onset in other regions of the whole-brain network (if and when present within the simulation period) is performed as follows. First, the baseline firing activity of each region is defined (first time window of the simulation). The non-SLE nodes initially exhibit a relatively stable firing rate, which gradually becomes influenced by the SLE originating from the Arbor node. When connected to the Arbor node, spiking-tonic brain areas are repeatedly recruited into SLE activity patterns, followed by periods of relaxation. For each region, we define the baseline firing rate as the minimum firing rate observed during the relaxation phase immediately preceding any event in the onset region.
When a robust increase in firing rate is detected in a region—indicative of a seizure-like event—the corresponding onset timestamp is recorded. The baseline firing rate serves as a reference, and seizure activity is identified based on a sustained increase in firing rate over a defined duration. We define the onset of a seizure as the beginning of the high-amplitude bursting regime, specifically the point at which the firing rate begins to rise consistently after the first small activity peak. To ensure that only significant deviations are classified as seizures, we require the firing rate to exceed the baseline by at least 10$\%$ and remain above this threshold for a minimum of 100 ms.
Our seizure onset detection follows a similar (though not identical) approach to that used in related studies—for example,~\citet{Melozzi_etal_2017,courson2024exploratory} — where seizure-like or bursting activity in mouse brain models is identified by monitoring a model variable and applying a threshold to detect the onset. In \textbf{Figure}~\ref{fig:seizure_prop_ca1}, we also depict the activity time-series of four initially non-SLE nodes of the brain network being recruited in the seizure, namely the ventral part of left Lateral Septal Nucleus (l~LSv), l~CA3, l~ENTCl and r~ENTCl. In the main panel (mouse brain template), non-colored regions correspond to areas where seizures either did not occur or had relatively weak effects.

The Arbor network was initialized with fixed connection weights ($w = 0.5$). To assess robustness, we also tested weights drawn from a normal distribution ($\mu = 0.5$, $\sigma = 0.5$), truncated to positive values. Neural activity was averaged over 20 independent realizations (see \textbf{Figures S2} and \textbf{S3} in the Supplemental Material). Experiments with inhibitory/excitatory network variants are also currently in progress.
Both baseline and seizure-evoked activity differ across brain regions. Due to the symmetrical inter-hemispheric connections in the Allen Mouse Brain SC, l~ENTCl and r~ENTCl share the same baseline firing rate. As the SLE pattern emerges in the left hippocampus, l~ENTCl exhibits higher spiking rates.

\subsection{Computational performance of the Arbor-TVB co-simulation framework}

We next evaluated the performance of the running example on a single Apple M1 (2021) laptop. Arbor was built with MPI and SIMD (Arm Neon/SVE) support, with cells organized into groups of ten to fully exploit SIMD capabilities. The overall runtime consists of four primary components: (1) Arbor model update, (2) Conversion from spikes to rates, (3) TVB model update, and (4) Conversion from rates to spikes.

\begin{figure}[htbp]
    \centering
    \includegraphics[width=\linewidth]{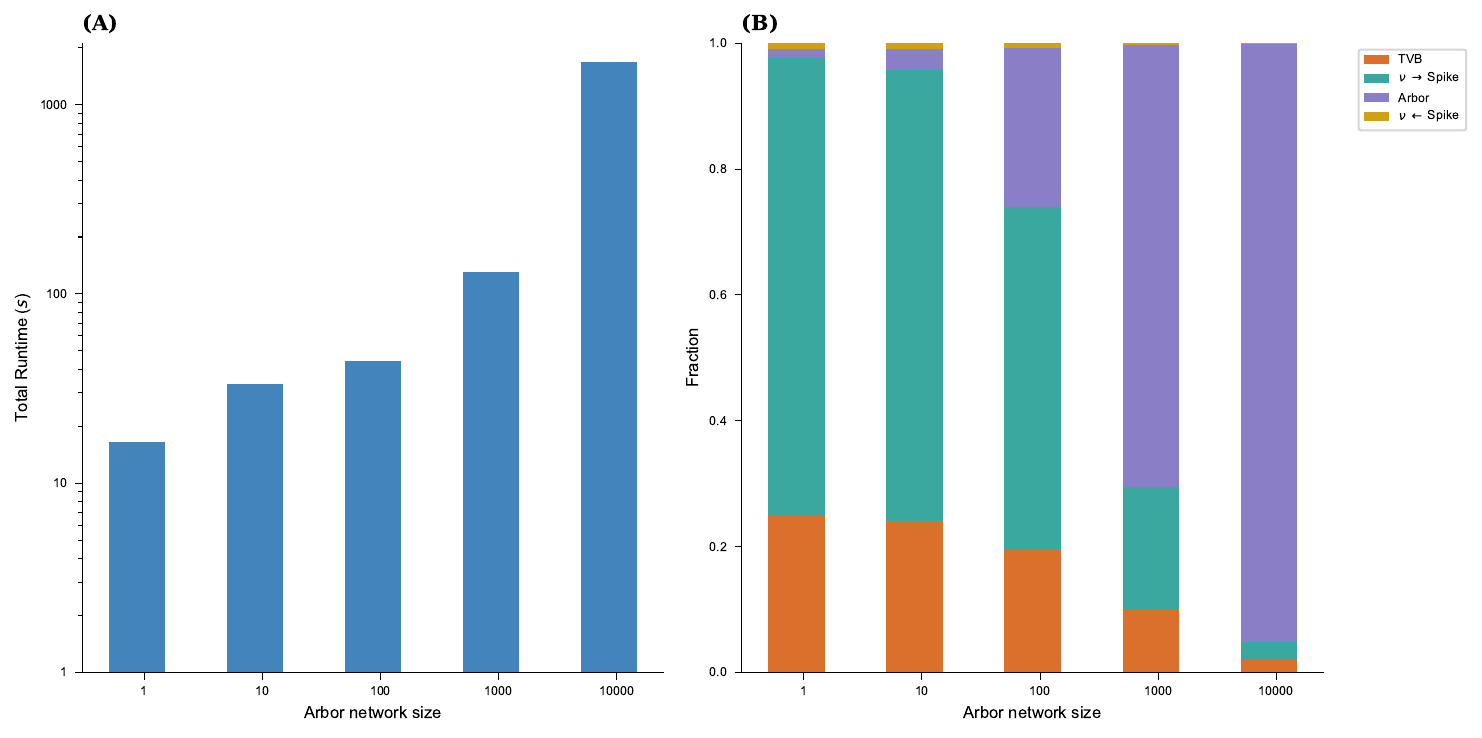}
    \caption{\textbf{Performance of the running example.}%
      We graph the overall runtime \textbf{(A)} of the full co-simulation over
      the number of cells in the Arbor network as well as the fraction of the
      main contributions \textbf{(B)}: time spent in both simulation engines and
      in converting between rates and spikes. TVB and rate to spike conversion
      are the most relevant cost center in the limit of vanishing Arbor network
      sizes while at large sizes, Arbor compute time dominates.}%
    \label{fig:cosim-perf}
\end{figure}

The Arbor update runs in parallel with the conversions between rates and spikes, as well as the TVB update. During spike exchange, both simulations synchronize, meaning the slower part must wait in the MPI collective, which accounts for the primary time spent in the collective call. \textbf{Figure}~\ref{fig:cosim-perf} illustrates the total runtime of a \qty{10}{s} simulation for the entire model described above, along with the relative contributions, for system sizes ranging from one to \num{10000} cells. Notably, at \num{10000} cells, nearly all computational time is spent within the Arbor network model. In future experiments, we plan to leverage additional hardware, including GPUs, to accelerate the Arbor side of the simulation. At this scale, TVB and the spike/rate conversions are potential bottlenecks that will require optimization, potentially through TVB’s JIT compilation and GPU acceleration. Additionally, further parallelization and porting of the conversion steps to a more performant programming environment remain promising avenues for improvement.

\section{Discussion}

In this work, we presented a co-simulation framework that offers a novel
approach to bridging the gap between microscopic (spiking neuron) and
macroscopic (mean-field) models. This framework integrates simulators Arbor and
TVB within a parallelized MPI environment, enabling a detailed yet
computationally feasible representation of neural dynamics across scales. To
model large-scale brain network dynamics, we used the mean-field reduced
Wong-Wang model to reproduce resting-state dynamics. Simultaneously, a detailed
spiking neural network was simulated with Arbor, employing a physiological model
of seizures at the neuron level (\textbf{Figure}~\ref{fig:single_neuron_kbath}).
The spiking activity of the population was then converted into a smooth trace
for the proxy node, which was communicated with TVB
[\textbf{Figure}~\ref{fig:tvb-model}(B)]. This co-simulation approach
successfully captured the interplay between spiking activity and large-scale
brain dynamics, where local neuron dynamics generate global activity
wave-fronts. At the microscopic scale, we demonstrated that the structure of the
detailed neural network influences its activity patterns, thereby affecting the
shape of the activity wavefront (\textbf{Figure}~\ref{fig:co-sim-example}).

As a proof of concept, and as a technical showcase, we simulated the emergence
of seizure-like activity patterns (SLE) in the mouse hippocampus, using the
Allen Mouse Brain Structural Connectivity data. By tuning the single neural
parameter, we modeled the target brain area with a small, fully-connected
network of (point and detailed-compartmental) neurons SLE, which produces
scale-free activity patterns. This network design can be further adapted to
investigate more bio-inspired scenarios, such as the study of: (i) ensembles of
neurons within a given brain area that belong to different sub-regions (i.e.,
are relatively distant from one another), whose dynamical activity may differ
significantly from that of other sub-groups and (ii) dynamical transition
phenomena where the $K_{\text{bath}}$ value is set near a critical
threshold—e.g., inducing a transition from regular spiking to bursting activity,
see \textbf{Figure}~\ref{fig:single_neuron_kbath}(A,B). The latter consideration
is more general, in the sense that for a different dynamical model of neural
activity, one can choose a relevant control parameter that drives transitions in
neural activity patterns—for example, from slow spiking (representing healthy
activity) to fast bursting (associated with pathological activity) in a given
region, when the system operates near a critical transition point (e.g., from
spike trains to bursting). Our approach offers a well-understood and easily
controlled platform by showcasing its technical usability. By adapting the Arbor
nodes with different cell composition and connectivity, we demonstrated its
significant potential for more complex cell models untapped.

Although using seizure propagation as a use case, the seizure activity patterns
established in the present study may not capture precisely the complex nature of
all epileptic seizures. For example, there are other types of seizures that
occur in different brain disorders, i.e., acute symptomatic seizures, which are
not like those caused in epilepsy. For example, patients with Alzheimer's
disease may also experience seizures, which are classified as progressive
symptomatic seizures and typically arise from underlying neurodegenerative
processes., see e.g.,~\citep{Mauritz_etal_2022} for a recent relevant review.

The implementation in the present study extended the functionality and
application scenarios of Arbor. Arbor has embarked on many types of
computational studies as a new-generation simulator. It enables seamless
conversion and simulation of single-neuron models from the NEURON simulator and
supports simulations of both individual neurons and large-scale networks. Arbor
accommodates various plasticity models, including spike-timing-dependent
plasticity, calcium-based synaptic tagging and capture, and structural
plasticity. It has been used to study synaptic tagging and capture via the
built-in diffusion functionality [\cite{luboeinski2024plasticarbor}, under
review]. Recent developments focus on co-simulation with membrane dynamics and
external kernels, enabling dynamic connectivity modifications in a
distance-dependent manner. With its high flexibility and scalability, Arbor
stands out as a promising platform developed within the EBRAINS initiative to
advance cross-scale simulations in computational neuroscience. Arbor is
available as part of the EBRAINS software distribution (ESD) on connected HPC
centres and the EBRAINS collab via Jupyter Lab. Providing a bridge between
morphologically detailed neurons and neural mass models encompassing the full
brain spans a gap of scale from sub-micrometer to decimeter. It allows for
placing the resolution---using Arbor and detailed models---where needed and
using realistic, data driven environments everywhere else via TVB.\@

Despite its successes, the Arbor-TVB framework has some limitations. The
co-simulation requires careful exploration and calibration of coupling
parameters to ensure meaningful interactions between Arbor and TVB, which
remains a challenge when generalizing to diverse neural models. The TVB network
we used here is homogeneous, in the sense that all model parameters are set to
be identical. While this is not a highly realistic assumption---particularly
when assigning a dynamical mean field model to simulate the activity of a brain
region---it is a common choice among researchers when modeling whole-brain
resting-state dynamics, see e.g.,~\citet{Popovych_2021,manos_etal_2023} and
references therein. Note that even with identical initial settings, firing rates
vary across regions due to the influence of long-distance connectivity weight
values and the resulting complex dynamics. Our use-case simulation of seizure
propagation is not yet directly compared to experimental data from mice or
humans. Nevertheless, the current Arbor-TVB implementation is capable of
indirectly capturing biologically realistic brain dynamics and activity
propagation, as reported for example in~\citet{Melozzi_etal_2017}
and~\citet{courson2024exploratory}. The same applies to the choice of calcium
kernel parameters. This kernel is inspired by experimental observations via
calcium imaging experiments~\citep{grewe2010high} but no exact values are
available. Due care should be taken when matching the magnitudes of both Arbor
output and TVB output. Moreover, within the Arbor-TVB framework, the chosen
dynamical model can be further tuned to generate neural activity — such as BOLD
signals or firing rates — that more closely aligns with neuroimaging data,
thereby enabling the simulation of more realistic brain dynamics. Specifically,
computational costs for large neural networks may necessitate further
optimization in model parallelization and data handling. A near-term goal would
be to incorporate new features, such as synaptic plasticity, which could offer
valuable insights into how brain networks adapt and reorganize in response to
disrupted activity.

As stated in the Introduction section, it is natural to compare the Arbor-TVB
co-simulation framework presented in the current Method paper with the
established NEST-TVB co-simulation~\cite{kusch2024multiscale}. Both simulators
have their own merits and application scope. NEST has a longer
history~\cite{Gewaltig:NEST} and a rich profile of neural models and plasticity
models. Its performance is excellent in mesoscopic modeling of spiking neural
networks. Arbor is younger but has made it possible to efficiently simulate
networks of the highly expensive biophysical HH model with
multi-compartments~\cite{arbor2019}. It also accommodates a variety of
plasticity rules, including heterosynaptic dendritic plasticity rules inside the
dendritic shaft~\cite{luboeinski2024plasticarbor}. The Arbor-TVB co-simulation
framework has thereby inherited those merits of Arbor, while the NEST-TVB
co-simulation framework opens the venue for users to freely use NEST functions
for co-simulation. Those two tools are complementary for distinct research
purposes. Moreover, since Arbor is designed to make the best use of both GPU and
CPU, Arbor-TVB can also be easily adapted to make use of the cutting-edge
exascale GPU computing resources.

From an epilepsy-seizure perspective, while the framework provides insights into
seizure propagation, additional validation against empirical data would enhance
its applicability to clinical settings. This co-simulation framework could
enable a detailed investigation of the physiological sources of seizures.
Understanding the impact of the structure of the diseased area on seizure
patterns and propagation would be of great interest [see e.g.,~\cite{Netoff2004,
  GarciaRamos2016}]. Specifically, we expect the inhibition and excitation
ratios in the detailed neural network to play a critical role in seizure
patterns [see e.g.,~\cite{Liu_2020, Engel_1996}]. Moreover, the Arbor-TVB user
can implement various neural models and configuration topologies to simulate
different brain regions and to computationally investigate diverse dynamic
activities or the effects of medical interventions—for example, modeling
subthalamic neurons along with synaptic and structural plasticity under
stimulation in Parkinson's disease~\citep{manos2021long,Meier2022,Shaheen2022},
Alzheimer's disease~\citep{Stefanovski_etal_2019,manos_etal_2023} or
tinnitus~\citep{Manos_2018a,Manos_2018b} etc. Evidently, comparison, parameter
tuning, and validation are also feasible using empirical neuroimaging data,
however this was not the primary goal of this work. The simulated neural
activity generated in each brain region of the connectome can be transformed
into a BOLD signal-similar to the built-in functionality of
TVB~\citep{sanz2013virtual,Melozzi_etal_2017}, which computes the hemodynamic
response function (HRF) kernel (i.e., fMRI activity) using the
Balloon–Windkessel model~\citep{Friston_2000}—and can ultimately be aligned with
neuroimaging time series data. Hence, a framework like Arbor-TVB can be extended
to investigate various brain conditions.

%%%%%%%%%%%%%%%%%%%%%%%%%%%%%%%%%%%%%%%%%%%%%%%%%%%%%%%%%%%%%%%%%
%%% End of Article

\section*{Conflict of Interest Statement}
%All financial, commercial or other relationships that might be perceived by the academic community as representing a potential conflict of interest must be disclosed. If no such relationship exists, authors will be asked to confirm the following statement: 

The authors declare that the research was conducted in the absence of any commercial or financial relationships that could be construed as a potential conflict of interest.

\section*{Author Contributions}

%The Author Contributions section is mandatory for all articles, including articles by sole authors. If an appropriate statement is not provided on submission, a standard one will be inserted during the production process. The Author Contributions statement must describe the contributions of individual authors referred to by their initials and, in doing so, all authors agree to be accountable for the content of the work. Please see  \href{https://www.frontiersin.org/about/policies-and-publication-ethics#AuthorshipAuthorResponsibilities}{here} for full authorship criteria.

TH:\@ Methodology, Investigation, Software, Formal analysis, Writing --- original draft, review and editing, Visualization. JC:\@ Methodology, Investigation, Formal analysis, Visualization, Writing --- original draft, review and editing. HL:\@ Methodology, Investigation, Visualization, Writing --- original draft, review and editing. SD:\@ Conceptualization, Methodology, Resources, Review and editing, Supervision. TM:\@ Conceptualization, Methodology, Resources, Writing --- original draft, review and editing, Supervision, Project administration, Funding acquisition.

\section*{Funding}
JC was supported by the LABEX MME-DII (ANR-16-IDEX-0008) PhD grant. This work was supported by the NIH 1R21AG087888-01 grant. HL is supported by EBRAINS2.0. EBRAINS 2.0 has received funding from the European Union's Research and Innovation Program Horizon Europe under Grant Agreement No. 101147319. Open access publication funded by the Deutsche Forschungsgemeinschaft (DFG, German Research Foundation) – 491111487.

\section*{Acknowledgments}
%This is a short text to acknowledge the contributions of specific colleagues, institutions, or agencies that aided the efforts of the authors.
The founders had no role in study design, data collection and analysis, decision to publish, or preparation of the manuscript.

\section*{Supplemental Data}
% \href{http://home.frontiersin.org/about/author-guidelines#SupplementaryMaterial}{Supplementary Material} should be uploaded separately on submission, if there are Supplementary Figures, please include the caption in the same file as the figure. LaTeX Supplementary Material templates can be found in the Frontiers LaTeX folder.
The Supplementary Material for this article can be found online at:\@ \dots (to be inserted by the journal)

\section*{Data Availability Statement}
%The datasets [GENERATED/ANALYZED] for this study can be found in the [NAME OF REPOSITORY] [LINK].
% Please see the availability of data guidelines for more information, at https://www.frontiersin.org/about/author-guidelines#AvailabilityofData
Upon acceptance, we will provide the full source code of the Arbr-TVB cosimulator and selected use cases under a permissive open-source license, which will be linked here and given a DOI.\@
In addition, parts of the results will be published as a tutorial in the Arbor documentation.

\bibliographystyle{Frontiers-Harvard} %  Many Frontiers journals use the Harvard referencing system (Author-date), to find the style and resources for the journal you are submitting to: https://zendesk.frontiersin.org/hc/en-us/articles/360017860337-Frontiers-Reference-Styles-by-Journal. For Humanities and Social Sciences articles please include page numbers in the in-text citations 
\bibliography{NETNbib}

\end{document}